\documentclass[aps,prb,onecolumn,superscriptaddress,floatfix]{revtex4}
\usepackage{epsfig,amsmath,amssymb,color}
\usepackage{graphicx}

\begin{document}

\title{Susceptibility at the edge points of magnetization plateau of 1D electron/spin systems}

\author{T. Vekua}
\affiliation{Institut f\"{u}r Theoretische Physik, Leibniz Universit\"at Hannover\\ 
Appelstra\ss e 2, 30167 Hannover, Germany}

\date{\today}
    
\begin{abstract}
We study the behavior of magnetization curve as a function of magnetic field in the immediate vicinity of 
the magnetization plateaus of 1D electron systems within the bosonization formalism. First we discuss the plateau 
that is formed at the saturation magnetization of 1D electron system. Interactions between electrons we treat 
in the lowest order of perturbation. We show that for isolated systems, where total number of electrons is not allowed to vary,
magnetic susceptibility stays always finite away of half filling. 
 Similar statement holds for many other magnetization plateaus supporting nonmagnetic gapless excitations 
encountered in 1D electron/spin systems in the absence of special symmetries or features responsible for the mode decoupling. 
We demonstrate it on example of the plateaus at irrational values of magnetization in doped  
modulated Hubbard chains. Finally we discuss the connection between the weak coupling description of saturation magnetization plateau and strong coupling description of zero magnetization plateau of attractively interacting electrons/ antiferromagnetically interacting spin 1 Bosons. 
\end{abstract}

\maketitle

\section{introduction}

Magnetization process of 1D electron systems is fairly well understood theoretically. For purely spin systems (where charge fluctuations are frozen)
the magnetization curve (magnetization plotted versus magnetic field) can show plateau like behavior (at zero temperature) for certain rational values of magnetization\cite{Oshikawa97} where magnetic excitations develop gap\cite{Totsuka98}.

Experimentally different materials have been sinthesized exhibiting plateaus in their magnetization curve, and having magnetic structure believed to be modeled
by quasi- one dimensional geometries. For example dimerized spin $S=1$ chain\cite{Narumi} exhibits magnetization plateau at $1/2$ of saturation magnetization as predicted by theory \cite{Totsuka98}. Other quasi-1D compounds showing plateau at half of the saturation magnetization are spin $S=1/2$ compounds\cite{Inagaki,Wada}. 

Phase transitions at the edges of the magnetization plateau is usually described within the commensurate- incommensurate universality class\cite{Japaridze, Pokrovski}. 
As a hallmark feature of commensurate - incommensurate phase transition the magnetization shows square root dependence on magnetic field in the vicinity to the edges of the plateau, giving rise to infinite magnetic susceptibility at the edge points. The simplest examples include: magnetization plateau of spin $S$ chain at the saturation magnetization and a plateau of integer spin $S$ antiferromagnetic Heisenberg chain (or even leg spin $S$ ladder) at the zero magnetization and we depict them on Fig.(1).

\begin{figure}
\label{integerspinplateau}
\begin{center}
  \includegraphics[scale=0.5]{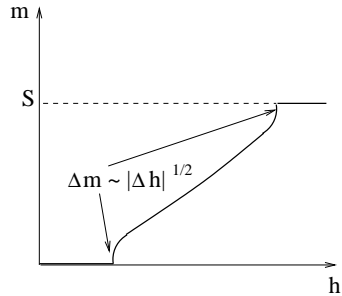}
\end{center}
\caption{magnetization curve of integer spin $S$ Heisenberg chain.
}
\end{figure}

When doped with charge carriers the plateaus can appear also at irrational (doping controlled) values of magnetization\cite{Cabra}. Interesting property of such plateaus is that they can support nonmagnetic gapless excitations. Other examples of plateaus with similar feature may be found in frustrated spin chains, like trimerized $S=1/2$ zig-zag chain at $1/3$ of saturation magnetization where gapless chiral excitations coexist with gapped magnetic excitations\cite{Rosales}, or spin $S$ generalization of the integrable $t-J$ chain doped with $S-1/2$ carriers\cite{Frahm}.

In this work we investigate in detail the edge points of plateaus with gapless nonmagnetic excitation where modes describing magnetic (gapped at the plateau) and non-magnetic (gapless) excitations do not decouple. We will argue that in such situations, in drastic contrast to the square root behavior characterizing the commensurate- incommensurate phase transitions, magnetization depends linearly on magnetic field at the edges of plateau, thus rendering magnetic susceptibility finite at the edge points.

The paper is organized in the following way: First we describe within the effective theory bosonization approach\cite{Tsvelik,Giamarchi} the simplest plateau of saturation magnetization encountered in free electron system. Next we add interactions between electrons and treat them in the first order of coupling constants. We compare our results to exact results available for the Hubbard model for both positive and negative Hubbard couplings. In the case of attractive Hubbard interaction we corrected the missprint in the exact expression of the saturation susceptibility\cite{Woynarovich86,Woynarovich91} and convinced that bosonization recovers exact expression of saturation magnetization in the lowest order of onsite Hubbard interaction for both signs. We apply the similar theory to other magnetization plateaus that are encountered in more complicated problems, supporting gapless excitations, and in particular to the plateau in dimerized doped Hubbard chain at doping $\delta=1-n$, where $n$ is a filling. As was shown in Ref.\cite{Cabra} at magnetization $m=\delta/2$ plateau is formed and we show that magnetization changes linearly with the field once plateau is closed by magnetic field on both sides as depicted on Fig.~(2).
Finally we discuss the connection between the weak coupling description of saturation magnetization plateau and strong
coupling description of zero magnetization plateau of attractively interacting electrons.

In appendix we give our bosonization rules and for completeness give exact expressions for saturation susceptibility for Hubbard model for both values of Hubbard coupling constant.

\begin{figure}
\label{dopedchainplateau}
\begin{center}
  \includegraphics[scale=0.5]{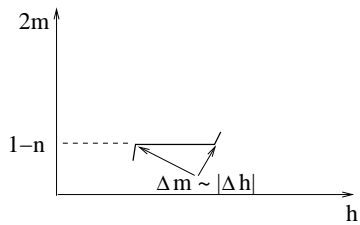}
\end{center}
\caption{Portion of the magnetization curve of a doped dimerized Hubbard chain.
}
\end{figure}

\section{Model and effective field theory}

The microscopic model, general enough for our purposes is described by the following lattice Hamiltonian of the one dimensional electrons:
\begin{eqnarray}
\label{latticemodel}
H&=&-\sum^N_{i,j=1}\sum_{\sigma}t_i(c_{i,\sigma}^{\dagger}c_{i+1,\sigma}+c_{i+1,\sigma}^{\dagger}c_{i,\sigma})+U\sum^N_{i=1}n_{i\uparrow}n_{i\downarrow} \nonumber\\
&&+V\sum^N_{i=1}\sum_{\sigma}n_{i\sigma}n_{i+1\sigma} -\frac{h}{2}\sum^N_{i=1}(n_{i,\uparrow}-n_{i,\downarrow})
\end{eqnarray}
We will always work at the fixed filling. Above $n_{i,\sigma}=c_{i,\sigma}^{\dagger}c_{i,\sigma}$ is the number operator of spin $\sigma=\{\uparrow, \downarrow\}$ electrons at the i-th site, $U$ is the onsite coupling constant, $V$ describes interaction between the same spin electrons on neighboring sites, $h$ is an external magnetic field.
$t_i=t+t'(-1)^i$ is modulated hopping amplitude with period 2 (dimerized).

We will study the continuum limit of the above model using bosonization approach\cite{Tsvelik,Giamarchi}. 
Bosonization rules are summarized in Appendix A.

In bosonic formulation effective field theory of free fermions (putting $t'=U=V=0$ in Eq. (\ref{latticemodel})) is just a direct sum of two decoupled Gaussian models, one for each spin component: 
\begin{equation}
{\cal H} =  \sum_{\sigma}\frac{v_{\sigma}}{2}\int dx \, \Big\{(\partial_x \phi_{\sigma})^{2} 
+ (\partial_x \theta_{\sigma})^{2}\Big\} \, .
\end{equation}
where velocities are determined by the fillings: $v_{\sigma}=2ta\sin{\pi n_{\sigma}}$, with $a$ standing for lattice constant that we will put to 1.

\section{Spin susceptibility at the saturation magnetization}
Magnetic susceptibility of the Hubbard model with fixed filling is always finite at the saturation magnetization for any interaction strength\cite{Woynarovich86,Woynarovich91,Carmelo} (including $U>0$, $U<0$, and $U=0$) except for half filling\cite{Takahashi69} and empty or completely filled bands. It is instructive first to develop bosonization analysis and reproduce the above result within the effective field theory language. This will serve to us as a warm up procedure for more complicated cases considered later.
In this section we put $t'=0$ and $t=1$ in Eq.~(\ref{latticemodel}).

\subsection{Bosonization for free electrons at saturation magnetization}

First we put $U=V=0$ in Eq.~(\ref{latticemodel}) and apply bosonization for free electrons at saturation magnetization.
At saturation magnetization band of down- spin electrons
gets empty, while the band of up- spin electrons gets either
partially filled, if filling is less than half, or completely filled exactly
at half filling. To mimic the bottom of the down spin electron band we add an appropriate relevant sine- Gordon potential to free bosonic theory describing the down- spin electrons (see Appendix B):
\begin{eqnarray}
\label{Hamiltonian}
H&=&\frac{v_{\downarrow}}{2}\left[(\partial_t\phi_{\downarrow})^2 +(\partial_x\phi_{\downarrow})^2 \right] +\frac{v_{\uparrow}}{2}\left[(\partial_t\phi_{\uparrow})^2 +(\partial_x\phi_{\uparrow})^2 \right]\nonumber\\
&+&\frac{1}{2}(h_{W}+\Delta h)\frac{\partial_x\phi_{\downarrow}}{\sqrt{\pi}}+W\cos{\sqrt 4\pi}\phi_{\downarrow}-\frac{\Delta h}{2}\frac{\partial_x\phi_{\uparrow}}{\sqrt{\pi}}\nonumber\\
&-&\Delta \mu(\frac{\partial_x\phi_{\downarrow}}{\sqrt{\pi}}+\frac{\partial_x\phi_{\uparrow}}{\sqrt{\pi}})
\end{eqnarray}
Above $v_{\uparrow}=2\sin(\pi n)$, is the velocity of up spin band at saturation, $h_W$ closes the gap opened by the cosine term and $\Delta h$ is a deviation of magnetic field from the saturation value. 
Down- spin electron parameters $v_{\downarrow}$ and $W$ are selected in such a way to mimic optimally the dispersion at the bottom of the down spin electron band. As it will become obvious we will not need their explicit values for calculating the magnetic susceptibility at the saturation magnetization.
The chemical potential is a function of magnetic field (Lagrange multiplier enforcing constraint) and keeps the number of electrons fixed.

In bosonization the constraint of fixed particle number looks as:
\begin{equation}
\label{constraint}
\left< \partial_x \phi_{\downarrow}+\partial_x\phi_{\uparrow}\right>=0
\end{equation}
On the other hand uniform magnetization density is given by the following expression:
\begin{equation}
\label{magnetization}
 \Delta m=  \frac{\left<\partial_x \phi_{\uparrow}-\partial_x\phi_{\downarrow}\right>}{2\sqrt {\pi}}
\end{equation}
At the saturation magnetization we have from Eq. (\ref{Hamiltonian}): 
\begin{equation}
\label{up}
\frac{\left< \partial_x \phi_{\uparrow}\right>}{\sqrt{\pi}}=\frac{\Delta h/2+\Delta \mu(h)}{\pi v_{\uparrow}}=\frac{\Delta h/2+\Delta \mu(h)}{2\pi \sin{\pi n}}
\end{equation}
Using Eqs.~(\ref{constraint}) and (\ref{magnetization}) from Eq.~(\ref{up}) we can express the Lagrange multiplier $\Delta \mu$ as follows:
\begin{equation}
\label{lagrange}
\Delta \mu=\pi v_{\uparrow}\Delta m-\Delta h/2
\end{equation}
In addition using the sine- Gordon Hamiltonian for down- spin electrons we have second equation\cite{Japaridze,Pokrovski} (for details see Appendix B):
\begin{eqnarray}
\label{satmagn}
-\Delta n_{\downarrow}&=&\Delta m=\frac{1}{\pi}\sqrt{|\Delta h/2-\Delta \mu|}\nonumber\\
&=&\frac{1}{\pi}\sqrt{|\Delta h/2-(\pi v_{\uparrow}\Delta m-\Delta h/2)|}
\end{eqnarray}
where for the last equality we have used Eq. (\ref{lagrange}).
The correct root of the Eq. (\ref{satmagn}), one vanishing with $\Delta h$, is:
\begin{equation}
\label{magn}
\Delta m=\frac {\Delta h} {\pi v_{\uparrow}}- \frac{(\Delta h)^2} {\pi v^3_{\uparrow}}+O(\Delta h^3)
\end{equation}

Yielding for the magnetic susceptibility at the saturation magnetization:
 \begin{equation}
\label{susc}
 \chi=\frac{\partial m}{\partial h}|_{h_{sat}}=\frac{1}{\pi v_{\uparrow}}=\frac{1}{2\pi \sin{\pi n}}.
\end{equation}

\subsection{Including interactions perturbatively}

From the Eq.~(\ref{susc}) we see that the susceptibility is finite except at half filling. Important observation is that this result is robust to interactions,  one can treat interactions between electrons {\it perturbatively}, because the saturation magnetization plateau is caused by the band effect and exists already at the free electron level. To demonstrate this we will include perturbatively Hubbard coupling $U$ and next nearest neighbor interaction $V$ and obtain exact expression of susceptibility in the first order of these coupling constants.  
Interactions $U$ and $V$ in bosonization formulation close to saturation magnetization are translated as:
\begin{equation}
\label{Hubbard}
\frac{U}{\pi}\partial_x\phi_{\downarrow}\partial_x\phi_{\uparrow}+\frac{V}{\pi}\partial_x\phi_{\downarrow}\partial_x\phi_{\downarrow}+\frac{V}{\pi}\partial_x\phi_{\uparrow}\partial_x\phi_{\uparrow}
\end{equation}
Other terms are either incommensurate- meaning that they rapidly oscillate and
cancel out in the continuum limit or irrelevant (like Umklapp
process at half filling for the attractive case) and thus are neglected. 

Using constraint (\ref{constraint}) from (\ref{magnetization}) one can express magnetization by $\partial_x\phi_{\uparrow}$ simply by minimizing the energy:
\begin{equation}
\left< \frac{\delta H}{\delta \partial_x\phi_{\uparrow}}\right>=0
\end{equation}
Where $H$ is given by the sum of (\ref{Hamiltonian}) and (\ref{Hubbard}).
This fixes magnetization in the ground state configuration as follows:
\begin{equation}
\Delta m=
\frac{\left<\partial_x\phi_{\uparrow}\right>}{\sqrt{\pi}}=\frac{\Delta
  h/2+\Delta \mu-(U-2V)\left< \partial_x \phi_{\downarrow}
  \right>/\sqrt{\pi}}{\pi v_\uparrow}
\end{equation}

\begin{equation}
\Delta m(1-\frac{U-2V}{\pi v_{\uparrow}})=\frac{\Delta h/2+\Delta \mu}{\pi v_\uparrow}
\end{equation}
 
\begin{equation}
\Delta \mu= \pi v_{\uparrow} \Delta m (1-\frac{U-2V}{\pi v_{\uparrow}}) -\Delta h/2
\end{equation}
Now we need another equation to exclude $\Delta \mu$. We adopt a mean field decomposition of
interaction terms and for the second equation we obtain:
\begin{eqnarray}
\Delta m&=&\frac{1}{\pi}\sqrt{|\Delta h/2 -\Delta \mu +(U-2V) \Delta m|}\nonumber\\
&=& \frac{1}{\pi}\sqrt{|\Delta h- \Delta m (\pi v_{\uparrow}-2(U-2V)) |  }
\end{eqnarray}

\begin{equation}
\Delta m=\frac {\Delta h} {\pi \tilde v_{\uparrow}}+O(\Delta h^2)
\end{equation}

where 
\begin{equation}
\tilde v_{\uparrow}= v_{\uparrow}\left(1- \frac{2(U-2V)}{\pi v_{\uparrow}}\right)
\end{equation}

For the susceptibility at the saturation magnetization in the first order of coupling constants we obtain:

\begin{equation}
\label{bosonizationsusc}
\chi(n)= \frac{1}{2\pi \sin(\pi n)}\left(1+ \frac{U-2V}{\pi \sin{\pi n}}\right)
\end{equation}
The plus and the minus signs in front of $U$ and $V$ respectively are intuitively clear: repulsive Hubbard interaction facilitates to fully polarize the system while repulsion between the similar spin components hinders it. Moreover number of nearest neighbors along the chain is twice the number of sites, hence contribution from $V$ is multiplied by factor of 2 in comparison with $U$. 

In the case of the pure Hubbard model ($V=0$) we corrected a misprint of factor of 2 in \cite{Woynarovich86,Woynarovich91} (one can repeat Bethe Ansatz calculation close to saturation magnetization and get convinced that bosonization indeed gives correct result in the first order of $U$ for both signs, see Appendix C).

The above effective theory developed at the edge of the saturation plateau is not only valid in this particular case, but also applies to other plateaus, e.g. those that are characterized by gapless non-magnetic excitations which we will discuss in the following section.

\section{Plateau of doped dimerized Hubbard chain at irrational value of magnetization}

In this section we will apply our bosonization description that we developed in previous sections 
 to calculate magnetic susceptibility at the edges of plateau of doped dimerized Hubbard chain ($V=0, t'\neq 0$) at doping dependent value of magnetizuation $2m=\delta$\cite{Cabra}.
This plateau shares the similar physics of the saturation magnetization one.  
The exact equivalence with the saturation plateau is due to the fact that description is perturbative in interaction parameter, plateau exists already at the level of quadratic in Fermi fields Hamiltonian (due to the modulation of exchanges) and the limit $U\to 0$ is non- singular.

The bosonized Hamiltonian at the plateau of dimerized doped Hubbard chain, (first setting Hubbard interaction to zero) is given by Eq. (\ref{Hamiltonian}), since at this plateau one degree of freedom (up- spin) remains gapless, whereas down- spin electrons meet the commensurability condition and thus are gapped. Now we can add perturbatively Hubbard interaction, or more general interactions, and follow the steps that we developed in previous section. In this way we will obtain susceptibilities in the lowest order in interaction constants.
Situation is manifestly similar to the saturation magnetization case. From this equivalence we can answer a question\cite{Cabra} on the critical properties at the edges of these plateaus. Namely, magnetization will increase always linearly with the magnetic field, unless the total number of electrons is allowed to vary. If no constraint on the total number of particle is imposed, plateau itself will disappear\cite{Cabra}. Once again, due to the linearly dispersive degree of freedom, potential Van Hove singularity from the bottom of the gapped mode is suppressed by the constraint of keeping total number of electrons fixed.

We think that similar effective theory can apply to the doping dependent magnetization plateau in an integrable spin -$S$ generalization of the $t- J$ chain doped with spin $S-1/2$ carriers\cite{Frahm}. As explained within the Bethe Ansatz method\cite{Frahm} in this model quantum numbers of excitations carrying spin and charge get coupled at the right edge of the doping dependent plateau and this can cause the linear increase of magnetization as opposed to the naively anticipated square root behavior.

\section{Spin susceptibility at the onset of magnetization}

More subtle situation arises when magnetization plateau is formed due to the interactions (four fermion involving terms), and to describe correctly the edge of the plateau one needs to use non- perturbative approach.  One example of such situation is the weak coupling limit of Hubbard model of attractively interacting electrons where the effects of curvature of the Fermi surface on commensurate- incommensurate transition in spin sector away of half filling were emphasized\cite{Vekua}. 
However in the dilute limit of the same model a simple approach was developed\cite{Vekua} using the reasoning along the lines suggested in \cite{Woynarovich91} which
allows to obtain exact expression for susceptibility at the onset of magnetization.

\subsection{Attractively interacting electrons}
In this section we remind the bosonization description of zero magnetization plateau of attractively interacting two component fermions\cite{Vekua}.
 In the dilute limit the similar approach as that at the saturation
 magnetization was developed, treating bound pairs and up spin electrons as independent
 particles, but constrained by the total number of particles being fixed. The
 independence could be justified by a dilute limit, and treating bound pairs
 as hard core bosons can be justified by the inherent strong coupling nature of the dilute limit. Thus in the dilute limit the 
heuristic model developed in strong coupling\cite{Vekua} is expected to yield exact expression for susceptibility.
For completeness we remind the calculations developed in Ref.[\onlinecite{Vekua}] in dilute limit. The Hamiltonian is given by the following expression:
\begin{eqnarray}
\label{heuristic}
H&=&\frac{v_{\uparrow}}{2}\left[(\partial_t\phi_{\uparrow})^2 +(\partial_x\phi_{\uparrow})^2 \right] +\frac{v_{p}}{2}\left[(\partial_t\phi_{p})^2 +(\partial_x\phi_{p})^2 \right]\nonumber\\
&-&\frac{(h_W+\Delta h)}{2}\frac{\partial_x\phi_{\uparrow}}{\sqrt{\pi}}+W\cos{\sqrt 4\pi}\phi_{\uparrow}\nonumber\\
&-& \mu(\frac{\partial_x\phi_{\uparrow}}{\sqrt{\pi}}+\frac{2\partial_x\phi_{p}}{\sqrt{\pi}}+n)
\end{eqnarray}
Where index p stands for the bound pairs. The above Hamiltonian is manifestly similar to
the one describing the effective theory at the saturation plateau except that magnetic field does not couple to the pairs. 
In the same way as outlined before one can easily work out the
magnetic susceptibility. The dispersion of uncompensated up spin electrons at the critical field  looks:
\begin{equation}
\label{quadraticinmomenta}
E_{\uparrow}(k)=\sqrt{v_{\uparrow}^2k^2+\Delta^2}-\Delta \simeq \frac{v_{\uparrow}^2k^2}{2\Delta}
\end{equation}
where $\Delta$ is a spin gap in the dilute limit, and $v_{\uparrow}$ is the velocity of uncompensated spins in the strong coupling (for Hubbard model, however their values do not appear in the expression of susceptibility. 
On the other hand bound pairs disperse linearly with velocity $v_p
\neq 0$ . Repeating analogous to the saturation case
calculations we obtain for magnetization:
\begin{equation}
\Delta m=\frac{1}{2\pi}\sqrt{\frac{\Delta} {v^2_{\uparrow}}  (  h-h_{cr} -v_p \pi \Delta m)}
\end{equation}
and for susceptibility:
\begin{equation}
\chi|_{n=const} = \frac{1}{\pi v_p}. 
\end{equation}
For Hubbard model $v_p=
\frac{2\pi n}{|U|}$ is the charge (soft) mode velocity, so we get: $\chi|_{n=const} =\frac{|U|}{2\pi^2 n}$. Thi is the low density limit of the exact strong coupling result: $\chi|_{n=const} =\frac{|U|}{2\pi^2 n (1-n)^2}$ \cite{Woynarovich91}.

Note in the picture of up- spin and bound pair particles 
$ \Delta m=\frac{1}{2} \Delta n_{\uparrow}=-\Delta n_{p}$ here, different from the case where we considered the up and down -spin electrons where we used $\Delta m=\Delta n_{\uparrow}$. The differing 1/2 factor comes from a simple
consideration sketched on Fig. (1) where one can see, that $\Delta N_{\uparrow,b}=2\Delta N_{\uparrow,a}$.

\begin{figure}
\label{twon}
\begin{center}
  \includegraphics[width=0.47\textwidth]{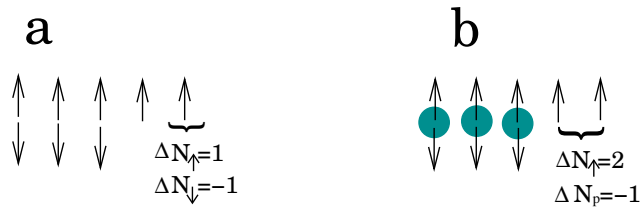}
\end{center}
\caption{Variation of up- spin electron number in two interpretations: a) up
  and down spin particles, and b) up spin particles and pairs. Both parts
  correspond to the same configuration of spins, but due to different
  interpretations up spin electron variations differ by factor of 2.}
\end{figure}

 This approach captures the intuitive physics at the commensurate- incommensurate
 phase transition: the spin susceptibility is diverging 
  because of the Van Hove singularity in the density of states of the
 uncompensated up- spin electrons when total number of electrons is not fixed. 
However, the linear dispersion of bound pairs, which also holds at
 the critical field, quenches the above-mentioned divergence when the total
 number of electrons is kept fixed\cite{Woynarovich91}. Certainly the above simple model provides
 with the correct physical picture, but it is valid in dilute limit, so it does
 not capture the special role of the half filled band. 
For the weak coupling limit close to half filling we refer to Ref.[\onlinecite{Vekua}].

\section{Summary}
We worked out effective field theory for the one dimensional electron systems 
describing the vicinity of the magnetization plateau. First we treated free electrons 
at saturation magnetization and then we added interactions in the lowest order of coupling constant. 
We showed that away of half filling (for bipartite models that we considered) magnetization
is a linear function of magnetic field at the edge of the saturation magnetization plateau if total number of electrons is
not allowed to vary. The same theory describes also other than saturation plateaus, e.g. we described one encountered 
in dimerized doped Hubbard chain. Similar reasoning is easily generalizable to other plateaus supporting gapless nonmagnetic excitations in 
polimerized doped Hubbard chains\cite{Cabra}. All these plateaus share the similar feature: they are adiabatically connected with non- interacting cases, that's why they are relatively easy to treat. More subtle situation arises when plateaus are opened due to the interaction, thus one can not describe them perturbatively. For such situations we explained in details the connection between the zero magnetization plateau of attractively interacting two component Fermi system in dilute limit\cite{Vekua} and saturation magnetization plateau of free electron system. 

The similar behavior of magnetization at the onset holds for spin 1 bosons\cite{remark}, where the spin degrees of freedom for zero magnetization are described by the $O(3)$ non linear Sigma model with gapped spectrum\cite{Zhou}, while charge degrees of freedom are gapless. 
Calculation simplifies in the dilute limit, which inherently falls under the strong coupling, where the system is made of strongly bound singlet pairs\cite{Cao,Essler} and the similar reasoning to that developed in Section V shows that magnetization increases linearly as soon as the magnetic field suppresses the spin gap if the total number of bosons is not allowed to vary.
However to calculate for general situation magnetic susceptibility at the edge of zero magnetization plateau one should go beyond the spin- charge decoupling approximation (which follows from linearized hydrodynamics) in the effective formulation.

Very generally when gapless excitations are present at the plateau and they are not decoupled from the gapped magnetic excitations due to special microscopic symmetries\cite{FrahmVekua}, magnetization of isolated system shows linear growth  with magnetic field instead of the square root behavior.

Our findings may be relevant for real quasi one-dimensional spin gapped materials, where one can observe experimentally crossover from square root dependence, at half filling, to linear growth of magnetization with magnetic field, away of half filling, in the vicinity of zero magnetization plateau.


\begin{acknowledgements} 

The work was started at LPTMS, a mixed research unit No. 8626 of CNRS and Universit\'e Paris Sud. Author acknowledges support from the cluster of excellence QUEST. The communications from referees are also acknowledged, in particular drawing of attention to spin 1 bosons.
\end{acknowledgements} 

\appendix

\section{Bosonization conventions}

Our bosonization
rules are based on the following correspondence between the fermionic and bosonic operators\cite{Essler}:
\begin{equation}
\frac{c_{i,\sigma}}{a_0}\to\frac{\eta_{\sigma}e^{ik_{F,\sigma}x }   }{\sqrt {2\pi} }e^{i\sqrt{4\pi}\Phi_{R,\sigma}(x)}  +\frac{\eta_{\sigma} e^{-ik_{F,\sigma}x } }{\sqrt {2\pi}}e^{-i\sqrt{4\pi}\Phi_{L,\sigma}(x)}
\end{equation}
 Where $k_{F,\sigma}=\pi n_{\sigma}$ are Fermi momenta expressed in terms of fillings: $n_{\sigma}=\frac{1}{N}<\sum^N_{i=1} n_{i,\sigma}>$,
$\eta_{\sigma}$ are Klein factors, ensuring fermionic commutation rules between electrons of different spin orientation:
\begin{equation}
\{\eta_{\sigma},\eta_{\sigma'}\}=2\delta_{\sigma,\sigma'}
\end{equation}

and $\Phi_{R,\sigma}$ and $\Phi_{L,\sigma}$ are right and left chiral bosonic fields of spin $\sigma$ electrons with the following equal time commutation relations:
\begin{eqnarray}
\label{commutations}
&&[\Phi_{R,\sigma}(x),\Phi_{L,\sigma'}(y)]=\delta_{\sigma,\sigma'}\frac{i}{ 4}  \nonumber\\
&&[\Phi_{R,\sigma}(x),\Phi_{R,\sigma'}(y)]=\delta_{\sigma,\sigma'}\frac{i}{4} sign(x-y) \nonumber\\
&&[\Phi_{L,\sigma}(x),\Phi_{L,\sigma'}(y)]=-\delta_{\sigma,\sigma'}\frac{i}{4}sign(x-y)
\end{eqnarray}

From these chiral fields we construct the usual bosonic fields, and their dual counterparts:
\begin{eqnarray}
\phi_{\sigma}(x)&=&\Phi_{L,\sigma}(x)+\Phi_{R,\sigma}(x)\nonumber\\
\theta_{\sigma}(x)&=&\Phi_{L,\sigma}(x)-\Phi_{R,\sigma}(x)
\end{eqnarray}
with non- local duality relation between them: 
\begin{equation}
[\phi_{\sigma}(x),\theta_{\sigma}(y)] = i\Theta (y-x)
\end{equation}
where $\Theta(x)$ is the Heaviside step function.
Electron number operators are expressed in bosonization as follows:
\begin{equation}
n_{i,\sigma} \to n_{\sigma}+\frac{1}{\sqrt{\pi}}\partial_x\phi_{\sigma}(x)+\frac{1}{\pi}\sin (2k_Fx+\sqrt{4\pi}\phi_{\sigma}(x))
\end{equation}

\section{Free electrons at saturation magnetization}

Here we give a brief review of the magnetization process of free electrons in the vicinity of saturation magnetization. For simplicity first we consider the grand canonical system, where the total number of electrons is not fixed. 
For external magnetic fields exceeding the saturation value, down- spin electrons band gets completely empty in the ground state. With decreasing the magnetic field from its saturation value the magnetization will also decrease. Change of magnetization will have two contributions: 1) from the partially filled up- spin electrons and 2) from the dilute down- spin electrons. The contribution from partially field up- spin electrons is linear with the field and it is given by Eq.~(\ref{up}) (where for grand canonical systems one should omit the Lagrange multiplier).
Let us discuss the contribution from the down- spin electrons. Close to the saturation the density of down spin electrons is vanishingly small. In the absence of interactions the ground state energy density, given by its kinetic part is:
\begin{equation}
\label{gse}
 e_{0\downarrow}=\frac{E_{kin\downarrow}}{L}=\frac{\pi^2 {n_{\downarrow}}^3}{3}, 
\end{equation}
where $n_{\downarrow}={N_{\downarrow}}/{L}=\Delta n_{\downarrow}$ and mass was normalized to 1/2 (we remind that $\hbar=1$ throughout the paper).
By definition chemical potential is given by the following expression:
$\mu_{\downarrow}= {\partial e_{0\downarrow}}/{\partial n_{\downarrow}}$
which using Eq. (\ref{gse}) gives: 
\begin{equation}
\label{downdensity}
n_{\downarrow} = \frac{\sqrt{ \mu_{\downarrow}}}{\pi}\end{equation} and accordingly:

\begin{equation}
\label{suscepvanhove}
\chi_{\downarrow}^{-1} = \left(\frac{\partial n_{\downarrow}}{\partial \mu_{\downarrow}}\right)^{-1}= 2\pi \sqrt{ \mu_{\downarrow}} 
\end{equation}
in particular we get that $\chi_{\downarrow}^{-1}\sim n_{\downarrow}\to 0$.

For free fermions $\mu(T=0)=E_F$, $N=L\int_0^{E_F} \chi\mathrm{d}\mu$, 
thus $\chi$ is nothing else than density of states and its divergence at the bottom of the band is termed the Van- Hove singularity in the density of states. Most importantly this divergence is a universal feature - this result is not restricted only to free fermions: interactions (short range, falling quicker than $1/x^2$) do not modify the above behavior. This one can check e.g. from exact solution of $XXZ$ spin chain. Indeed in $XXZ$ model, close to saturation, magnetization depends as a square root on deviation of magnetic field from the saturation value and numerical prefactor in front of the square root is independent of the value of magnetic anisotropy (Z part of the exchange).

The relation between the particle density and chemical potential, given by Eq.~ (\ref{downdensity}), holds also for massive relativistic fermions (Luther- Emery model), when chemical potential slightly exceeds the spectral gap, since the only ingredient needed for the above relation is the low energy dispersion, which must be quadratic in momenta - Eq.~ (\ref{quadraticinmomenta}). 
The Luther- Emery model on the other hand is equivalent to the sine-Gordon model, that is a model of real scalar field selfinteracting  with $\cos\beta \phi$ term for particular value of $\beta=\sqrt{4\pi}$\cite{Tsvelik,Giamarchi}. This explains origin of Eq.~ (\ref{satmagn}) which follows from Eq.~ (\ref{downdensity}), where for chemical potential one has to put: $\Delta h/2-\Delta \mu$, which couples to the density of down spin electrons given in bosonization by: $\partial _x \phi_{\downarrow}/\sqrt{\pi}$. 

Now we discuss how the situation changes when the total number of electrons is kept fixed while changing the magnetic field across the saturation value.
One can work out easily for free electrons that in order to keep the total number of electrons fixed while changing the magnetic field across the saturation value ( $h=h_{sat}\to h_{sat}-\Delta h$) one has to adjust chemical potential (serving as Lagrange multiplier) accordingly: 
\begin{eqnarray}
\label{chemi}
\Delta \mu(h) 
&=&\frac{1}{2}(4(1-\cos\frac{\pi n}{2})-\Delta h)\nonumber\\
&+&  2\cos\left(\frac{\pi n}{2}+\arcsin \frac{4\sin^2\frac{\pi n}{2}-\Delta h}{4\sin\frac{\pi n}{2}}\right)
\end{eqnarray}

One observes that for all fillings, except for half filling (note that at half filling $\Delta \mu(h)=0$):
\begin{equation}
\label{allfillings}
\frac{\partial \mu(h)}{\partial h}|_{h=h_{sat}}=\frac{1}{2}
\end{equation}

Calculating the second derivative from (\ref{chemi}) we obtain:
\begin{equation}
\label{secondorder}
\frac{\partial^2\mu(h)}{\partial h^2}|_{h=h_{sat}}=-\frac{1}{2\sin^2{\pi n}}
\end{equation}
From Hamiltonian (\ref{Hamiltonian}) we get for the down- spin electrons density:
\begin{eqnarray}
&&<\Delta n_{\downarrow}>=-\Delta m=-\frac{1}{\pi}\sqrt{|\Delta \mu-\Delta h/2|}\\
&&=-\frac{1}{\pi}\sqrt{ \left| \frac{\partial \mu(h)}{\partial h}|_{h_{sat}}\Delta h +\frac{1}{2}\frac{\partial^2\mu(h)}{\partial h^2}|_{h_{sat}}(\Delta h)^2-\Delta h/2\right|}\nonumber
\end{eqnarray}
Using Eqs.~(\ref{allfillings}) and (\ref{secondorder}) we obtain:
\begin{equation}
<\Delta n_{\downarrow}>=-\frac{\Delta h}{2\pi \sin{\pi n}}=-\Delta m
\end{equation}
Thus recovering the anticipated result.

\section{Exact expression for saturation susceptibilities}
Here for completeness we write out the Bethe Ansatz results of the susceptibilities for repulsive\cite{Carmelo,Essler} and attractive\cite{Woynarovich86,Woynarovich91} Hubbard interactions at the saturation magnetization (with the correcting the misprint for attractive case). Close to the saturation magnetization ground state energy of the Hubbard model, being an analytic function can be expanded in the Taylor series in infinitesimal deviation of the magnetization from its saturation value $\Delta m$:
\begin{equation}
e_0(\frac{n}{2}-\Delta m)=e_{FM}-h_{sat}\Delta m+\frac{\chi^{-1}_{sat}}{2}(\Delta m)^2+\cdots
\end{equation}
where saturation field is: $h_{sat}=\frac{\partial e_0}{\partial m}|_{m=n/2} =
\lim_{\Delta m \to 0}\frac{e_{FM}-e(\frac{n}{2}-\Delta m)}{\Delta m} $
and the inverse saturation susceptibility is: $ \chi^{-1}_{sat}= \frac{\partial^2 e_0}{\partial m^2}|_{m=n/2}$.
From the above expansion it follows, that in case
$\chi^{-1}\neq 0$,  $\Delta m \sim \Delta h$ and in case $\chi^{-1}= 0\to \Delta m \sim \sqrt{\Delta h}$. The most general form allowed for the systems where ability to expand ground state energy into the Taylor series in magnetization can be proved is $\Delta m \sim (\Delta h)^{1/n}$, with $n$ being positive integer.

Repeating the calculation presented in details in
\cite{Woynarovich86} and correcting the missprint
one obtains expressions of saturation susceptibilities:

\begin{eqnarray}
\chi^{-1}(U<0)|_{n}&=&\frac{2}{\pi} \sin(\pi n)\left[2\pi-2\arctan\left(\frac{4\sin(\pi n)}{|U|}\right)\right]^2\nonumber\\
\chi^{-1}(U>0)|_{n}&=&\frac{2}{\pi} \sin(\pi n)\left[2\arctan\left(\frac{4\sin(\pi n)}{U}\right)\right]^2
\end{eqnarray} 

It is easy to verify that these expressions to first order of $U$ indeed recover the bosonization result given by Eq.~(\ref{bosonizationsusc}). We note as well that in the first order of $U$ susceptibilities for both signs of $U$ have identical form, in accordance with the bosonization, this way the typo was identified in the case of attractive interaction\cite{Woynarovich86,Woynarovich91}.


\begin{thebibliography}{10}
\bibitem{remark} In general similar scenario should apply for any spin $S$ Bose or Fermi gases with gapped spin and gapless charge modes.
\end{thebibliography}
\end{document}